\documentclass{article}

\usepackage{arxiv}

\usepackage{hyperref}       
\usepackage{url}            
\usepackage{booktabs}       
\usepackage{amsfonts}       
\usepackage{nicefrac}       
\usepackage{microtype}      
\usepackage{lipsum}
\usepackage{graphicx}       
\usepackage{graphics}
\usepackage{cite}
\usepackage{subfig}

\usepackage{dblfloatfix}

\usepackage{enumitem}
\usepackage{amsmath}
\usepackage{amsfonts}
\usepackage{gensymb}
\usepackage{amssymb}
\usepackage{amsmath}
\usepackage{tikz}
\usepackage{pgfplots}
\usepackage{textcomp}
\usepackage{multirow}
\usepackage{siunitx}
\usepackage[mathscr]{euscript}
\usepackage{xspace}
\usepackage{enumerate}
\usepackage{hyperref}
\usepackage{url}
\usepackage{multirow}
\usepackage{tikz}
\usepackage{pgfplots}
\usepackage{textcomp}
\usepackage[utf8]{inputenc}
\usepackage{siunitx}

\title{A COMPREHENSIVE VISION ON CLOUD COMPUTING ENVIRONMENT: EMERGING CHALLENGES AND FUTURE RESEARCH DIRECTIONS 
}

\author{ Sakshi Chhabra$^a$, Ashutosh Kumar Singh$^b$ \\
$^a$Panipat Institute of Engineering and Technology, India \\
$^b$Department of Computer Applications, National Institute of Technology,\\
    Kurukshetra,  India\\
    $sakshichhabra555@gmail.com$, $ashutosh@nitkkr.ac.in$
}

\begin{document}
\maketitle

\begin{abstract}
Cloud computing has become the backbone of the computing industry and offers subscription-based on-demand services. Through virtualization, which produces a virtual instance of a computer system running in an abstracted hardware layer, it has made it possible for us to share resources among many users. Contrary to early distributed computing models, it guarantees limitless computing resources through its expansive cloud datacenters, and it has been immensely popular in recent years because to its constantly expanding infrastructure, user base, and hosted data volume. These datacenter’s enormous and sophisticated workloads present a number of problems, including issues with resource use, power consumption, scalability, operational expense, security and many more. In this context, the article provides a comprehensive overview of the most prevalent problems with sharing and communication in organizations of all sizes. There is discussion of a taxonomy of security, load balancing, and the main difficulties encountered in protecting sensitive data. A complete examination of current state-of-the-art contributions, resource management analysis methodologies, load balancing solutions, empowering heuristics, and multi-objective learning-based approaches are required. In its final section, the paper examines, reviews, and suggests future research directions in the area of secure VM placements.
\end{abstract}

\keywords{Cloud Computing \and Resource utilization \and Security \and Load balancing \and VM placement}

\section{Introduction}
Over the past few years, cloud computing has experienced significant growth and has made numerous noteworthy contributions, making it a huge success. Because of its low cost, incredible performance, on-demand access, authentication, and other other opulent features, it emerged as a large-scale distributed computing paradigm \cite{1} \cite{2}. Technology related to cloud computing has been viewed as the next-generation architecture for IT solutions. In contrast to conventional systems, it allows users to move their data and application software to the network \cite{3} \cite{4}. The wide facilities supplied by cloud infrastructure, such as processing power, storage space, networks, and other computational resources, can be used by clients to set up and operate their own software, including operating systems and applications. Although the client does not manage or control the cloud infrastructure, they do so for the operating systems, storage, and applications, as well as maybe for the components they select \cite{5} \cite{6} \cite{7}\cite{8}\cite{9}.
Data sharing in the cloud, conveniently provides a flexibility to store a large amount of variable data, ensuring the emergence of a promising and fortunate technique for the users. However, the growth rate of useful static and dynamic data opportunities often initiates a necessity of analyzing the authorization ability for the user access convenience \cite{10} \cite{11}. Sharing the data with the third party also invites intruders and prying eyes who may alter or modify the data according to their observance. This leads to getting loopholes in the user accessibility validation and affects information utility conventions \cite{12} \cite{13}.
Today, being an emerging strategy along with the quick development of the internet industry, market size, and technology standards globally, cloud computing has become a trend. The data sharing tendency provides us many opportunities and benefits, but sometimes it may become a way for intruders to alter, leak or regenerate private data which is why data owners and other parties over cloud demand high level of security which can be conducted by practical effective schemes based on security and privacy preserving algorithms \cite{14} \cite{15} \cite{16}.

\textbf{Why we adopt Cloud Computing?}
\begin{itemize}
    \item 
	We can easily customize according to our needs: hardware, software, storage, network bandwidth, 
           speed etc..
    \item   	Flexibility in nature.  
    \item  High Utilization and High Availability.
    \item 	We can easily control the whole cloud with the help of software only - add/cancel/rebuild the
           resources instantly.
           \item We can scale up or down, small or large the resources according to our needs.
           \item 	Cost effective in nature, it is really efficient for Small/Medium size companies because no upfront 
	cost involved.
	\item Maintenance is too low, it reduces the size of a client’s IT department.
\end{itemize} 
\subsection{Motivation}
Because of the virtualization and multi-tenancy of the cloud, cloud systems currently face special security and load balancing concerns. 82 percent of businesses reported financial gains from their most recent cloud adoption initiative. This computing helps organizations go green since, according to 64\% of enterprises, using the cloud has helped them use less energy and produce less trash \cite{17} \cite{18} \cite{19}. A new security is necessary to allow users to access cloud resources effectively without disclosing their private information and sensitive data to unauthorised parties, such as cloud providers. Additionally, data owners have complete control over the security and privacy of their outsourced data and can at any moment check security parameters like data integrity \cite{20} \cite{21} \cite{22} \cite{23}.
Proper load balancing can assist in making the most use of the resources that are already available, which will improve the system's overall performance while lowering resource usage. Additionally, it aids in the implementation of failover, makes scaling possible, prevents bottlenecks and overprovisioning, speeds up response times, etc. At the same time, consumers are demanding more functionality at lower cost and less response time. One challenge is to minimize the cost of power consumption for which manufacturers are developing new methodologies, hindering the development of standard balancing methods and test equipment that could potentially lower power costs with more functionality. With efficient load balancing, it is possible to reduce resource usage for yet another good reason, making businesses greener and saving money \cite{24} \cite{25}.
\textbf{Motivation for using cloud computing is five-fold:}\\
\begin{itemize}
    \item 	33 percent of businesses used the cloud for information access from any device rather than to cut costs.
    \item 	82 percent of businesses reported financial gains from their most recent cloud adoption initiative.
    \item 	Cloud computing facilitates virtualization because cloud-based IT infrastructure can be geographically separated and virtualized, freeing startups from having to take the physical location of their data centers and IT infrastructure into account when making decisions about how to run their businesses.
    \item 64\% of businesses claim that implementing the cloud has reduced their waste and energy usage.
    \item Small businesses report less employee reluctance to adopting the cloud, and research shows that 52 percent of them have seen an improvement in data center utilization and efficiency, while 47 percent of them have seen a reduction in operating costs \cite{26} \cite{27}.
\end{itemize}

Being a new strategy, cloud is now getting attention as a way to lower capital expenditure and boost system effectiveness. It keeps user data on a sizable virtual storage system made up of several servers connected by a network \cite{28} \cite{29}.
\section{Top Challenges of Cloud Computing}
\begin{itemize}
    \item \textbf{Security \& Privacy}: Security is the main issue that everyone seems to agree poses a barrier with cloud computing. Nearly all surveys place data security and privacy issues at the top. Due to the frequent outsourcing of crucial services to outside parties, maintaining data privacy and integrity, ensuring data availability, and proving compliance are all more difficult with cloud computing \cite{30} \cite{38}.
    \\
    \item \textbf{Load Balancing}: It is a significant problem with cloud computing. It is a method that evenly distributes the dynamic local workload over all cloud nodes. This will stop nodes from being overloaded on some while being inactive or performing minimal work on others. It aids in maximizing resource usage and user happiness \cite{31} \cite{32}.
    \\
\item \textbf{	Resource Management}: The over provisioning and under provisioning types of difficulty occurs in cloud environment because of cloud resources are not utilizing properly [13]. Resource management is needed at various levels like hardware, software, virtualization etc. Job scheduling is a type of resource provisioning where jobs are executed in a particular order to optimize the parameters like execution time, execution cost, energy consumption, throughput etc [14]. In cloud computing, resource provisioning and resource scheduling are the two fundamental phases that make up resource management. The process of enabling virtualized resources for user allocation is called cloud resource provisioning. When a cloud service provider accepts a user's request for resources, it builds the right number of virtual machines and distributes them among users in accordance with their needs \cite{33} \cite{34}.
\\
\item \textbf{Service Quality}: One of the main reasons given by businesses for not transferring their business apps to the cloud is service quality. The availability, performance, and scalability criteria for running a production application on the cloud, in particular, are not sufficiently guaranteed by the SLAs offered by the cloud providers, in their opinion. Most of the time, businesses receive a return for the time the service was unavailable, although the majority of the current SLAs cover any financial losses. Enterprises won't host their mission-critical infrastructure in the cloud if sufficient service quality guarantees aren't made \cite{35} \cite{36}.
\\
\item 	\textbf{Migration issues}: If data migration from a system to the cloud isn't managed carefully, there could be serious concerns involved. To fix this, a migration strategy that works effectively with the existing IT infrastructure must be created. Companies frequently struggle to choose the best service model for their operations and instead choose to work with suppliers who enable the building of specialized computer environments \cite{37} \cite{39}.
\end{itemize}
\section{Security}
Although there are many advantages to data storage in remote locations, there is always a chance that confidential information will be changed, leaked, or even regenerated. The main security issues include confirming the eligibility of users, insider threats or vulnerabilities, malware attacks, external connectivity to the corporate network, a lack of qualified security employees, and many more difficulties. Therefore, as security is the part of computing that is given the most priority, it should come as no surprise that security concerns are quite important in the cloud environment. Identity management and authentication are essential in cloud computing since the cloud computing technique may involve storing sensitive user data on both client computers and cloud servers \cite{40} \cite{41}. The general view of the secure cloud as shown in Figure 1.
\begin{figure} [htbp]
	\centering
	\includegraphics[clip,trim=0cm 0cm 0cm 0cm, width=0.32\linewidth, height=6.5cm]{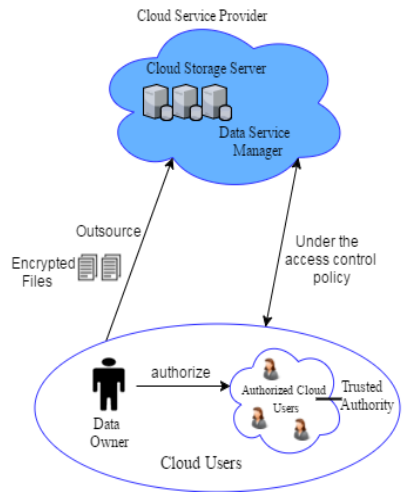}
	
	\caption{A General View of Secure Cloud}
	\label{Load Balancing Categorization1}
\end{figure}
There are a number of security risks associated with cloud computing that must be adequately addressed : 
\begin{itemize}
    \item \textbf{ Loss of governance}: Cloud service agreements might not contain a promise from the cloud provider to address these problems, leaving security defences vulnerable \cite{42} \cite{43}.
    \\
\item  \textbf{Authentication and Authorization}: Since sensitive cloud resources can be accessed from anywhere online, it is more important than ever to confirm a user's identity, especially since users may now include staff members, independent contractors, business partners, and clients. A crucial challenge is the need for reliable authentication and permission \cite{44} \cite{45} \cite{46}.
\\
\item \textbf{Failure of isolation}: The distinguishing characteristics of public cloud computing include shared resources and multi-tenancy. The failure of the systems that divide the use of storage, memory, routing, and even reputation by tenants falls under this risk category \cite{47} \cite{48}.
\\
\item	\textbf{Legal and compliance risks}: The cloud customer's investment in obtaining certification (e.g., to demonstrate compliance with industry standards or regulatory requirements) may be lost if the cloud provider is unable to provide evidence of its own compliance with the relevant requirements or refuses to permit audits by the cloud customer.
\\
\item \textbf{ Handling of security incidents}: Security breaches may be detected, reported, and then managed by the cloud provider, but customers still suffer as a result of these instances. To ensure that clients are not taken by surprise or notified with an unreasonable delay, notification requirements need to be specified in the cloud service agreement.
\\
\item  \textbf{Application Protection}: Applications have historically been protected by defense-in-depth security techniques based on a clear division of physical and virtual resources and trusted zones. Enterprises must reevaluate network-level perimeter security when responsibility for infrastructure security is passed to the cloud provider and install more controls at the user, application, and data levels \cite{49}.
\\
\item  \textbf{Data protection}: Here, sensitive data exposure or release as well as data loss or non-availability are the main worries. It could be challenging for the user of a cloud service to adequately audit the cloud provider's data handling procedures in their capacity as the data controller.
\\
\item  \textbf{Malicious behavior of insiders}: Given their access and privileges, employees who commit malevolent acts within an organisation have the potential to cause significant harm. This is made worse by the fact that such conduct may take place within either the customer organisation or the supplier organisation, or even both, in the cloud computing environment.\\
\end{itemize}
The data owners require high levels of protection and effective security measures from the cloud servers to mitigate these risks. Several efficient techniques and useful security algorithms can be used to provide this security. So that cloud servers can be utilised with confidence, our goal is to make the cloud secure and reduce these security vulnerabilities.\\
According to IDC, one of the top listed open concerns with implementing the cloud computing architecture is security. Although a lot of research into cloud security service engineering has been undertaken, most efforts and investigations has been done to achieve the fully secure environment. 

\section{Load Balancing}
Across a server cluster, a load balancer distributes network or application traffic as illustrated in Figure 2. Application responsiveness and availability are both enhanced by load balancing. Between the client and the server farm, a load balancer receives incoming network and application traffic and disperses it using a variety of strategies over numerous backend servers. A load balancer lessens individual server load by distributing application requests among several servers and avoids any one application server from becoming a single point of failure, hence enhancing the overall availability and responsiveness of the application. The simplest way to expand an application server infrastructure is through load balancing. New servers can be readily added to the resource pool as application demand rises, and the load balancer will start delivering traffic to the new server right away \cite{50} \cite{51}.
\begin{figure} [htbp]
	\centering
	\includegraphics[clip,trim=0cm 0cm 0cm 0cm, width=0.82\linewidth, height=5.5cm]{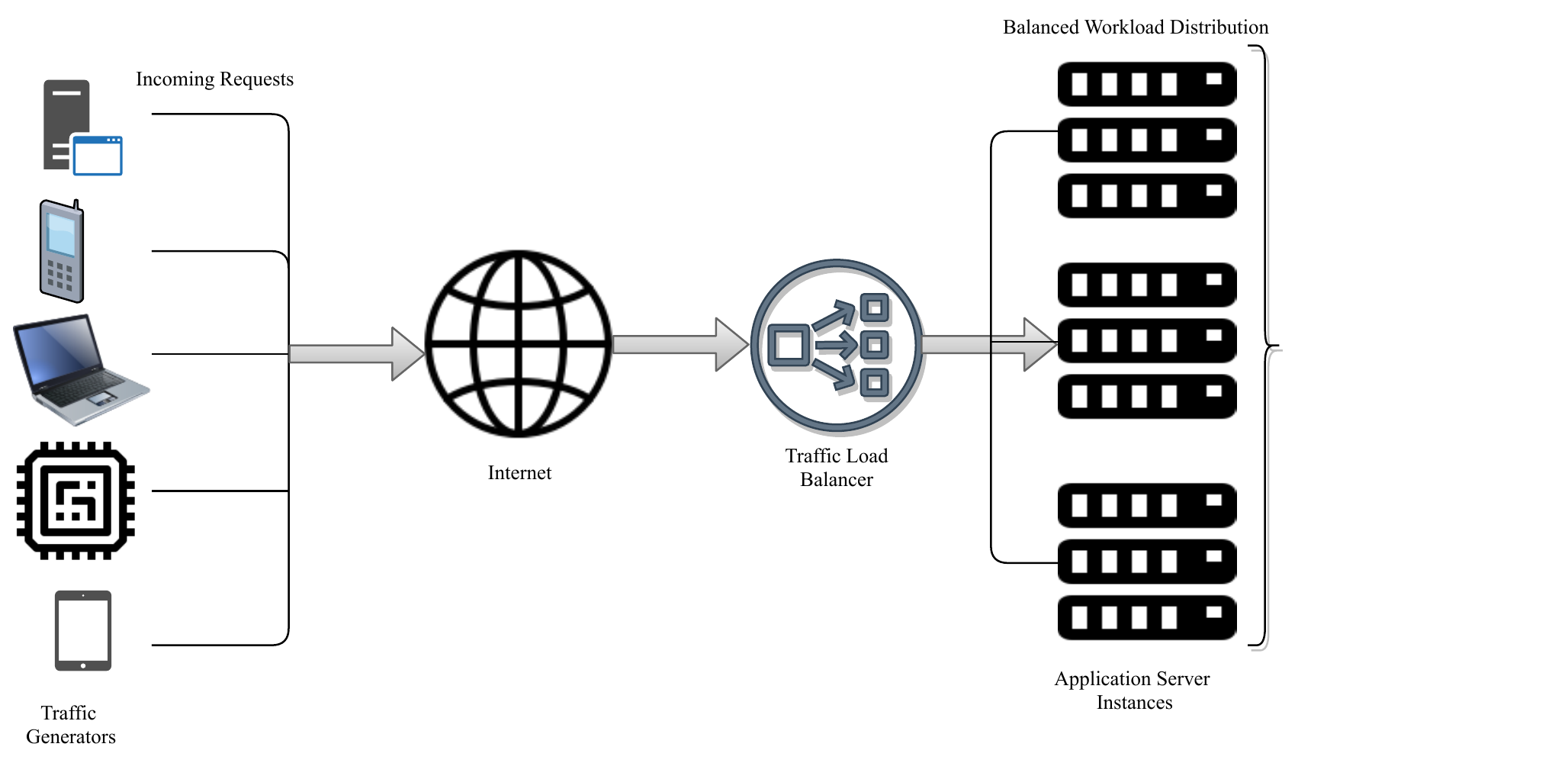}
	
	\caption{Load Balancing}
	\label{Load Balancing Categorization2}
\end{figure}
It is necessary to use load balancing to evenly distribute the tasks across the available resources in order to speed up computation and reduce task execution time. In general, some nodes may be fully utilized while others may be idle or underutilized. Therefore, better load balancing algorithms are able to improve performance, maintain system stability, build fault tolerance system, and accommodate future modification in addition to preventing from these scenarios \cite{52} \cite{53} \cite{54} \cite{55}. \\
Due to the importance and complexity of load balancing within the Cloud system, it is possible to intelligently load balance client access requests among server pools using a variety of methods and algorithms. Here, we discuss existing load balancing techniques:

\begin{itemize}
\item \textbf{Round-Robin Algorithm}: It is the static load balancing algorithm that distributes jobs using a round robin method. It chooses the first node at random and then distributes jobs evenly across the remaining nodes. The jobs are distributed to the processors in a circular fashion without regard for priority \cite{56}.
\\
    \item \textbf{Opportunistic Load Balancing Algorithm}: This load balancing mechanism is static, it does not take the VM's current workload into account. It tries to keep every node active. This technique swiftly completes jobs in a random sequence to the node that is now open.
    \\
    \item \textbf{Min-Min Load Balancing Algorithm}: In this kind of algorithm, the cloud manager first handles the jobs with the shortest execution times by allocating them to processors based on their capacity to finish the job within the allotted completion time. Jobs with long execution times must wait for an arbitrary amount of time.
    \\
    \item\textbf{ Max-Min Load Balancing Algorithm}: The Min-Min algorithm is used, with the exception that the cloud manager deals with jobs with the longest execution times after determining the minimal execution time.
\end{itemize}
\begin{figure} [htbp]
	\centering
	\includegraphics[clip,trim=0cm 0cm 0cm 0cm, width=0.82\linewidth, height=5.5cm]{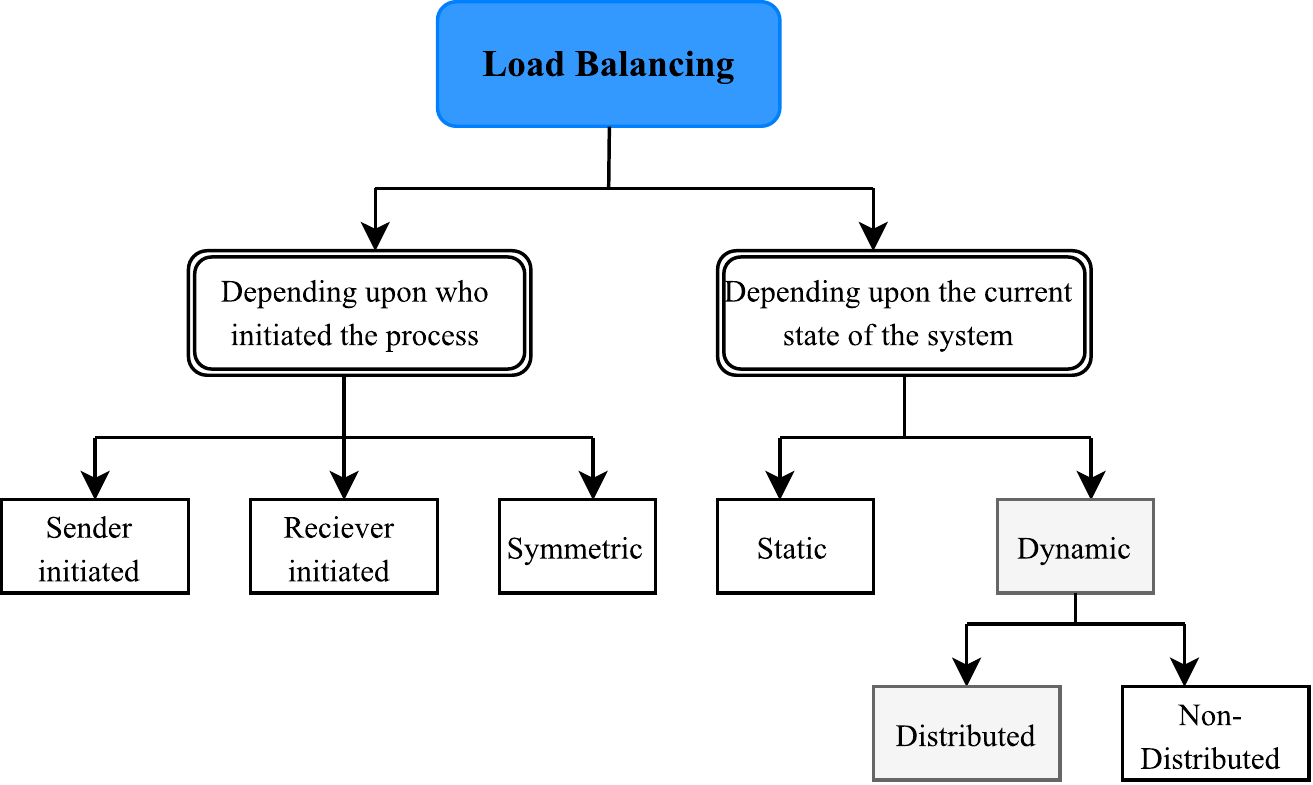}
	
	\caption{Load Balancing Categorization}
	\label{Load Balancing Categorization3}
\end{figure}
\subsection{Load Balancing Approaches}
Load Balancing have become an integral part of cloud resource management systems. A large body of early works focused on increasing scalability, edibility, performance and reduced downtime. Basically, load balancing is divided into two types: Depending upon who initiated the proces and depending upon the current state of the system can be seen in Figure 3. In recent years, researchers have started to look at how to maximize resource utilization like CPU, Disk, Memory, Power consumption etc. productively. The main approach is conducted in dynamic environment and ensures every computing resources that handle the tasks effectively \cite{34}. These techniques are applied to balance the future demands of resources \cite{70} \cite{71}, rate of SLA violation, the application performance and the execution time of jobs etc. Additionally, resource allocation, data preparation, and feature extraction from historical databases for training purposes all make use of machine learning techniques. For many of these methods, training is necessary to understand how applications behave \cite{57} \cite{58}.\\
\\
\textbf{Based on Heuristic Deployment}\\
\\
Using this approach \cite{59} \cite{60}, it is possible to select the best host for deploying the requested tasks in order to accomplish immediate load balancing. The working of the heuristic algorithms can be seen by this given flowchart in Figure 4. In order to acquire the ideal clustering set of physical hosts, they integrated the approach with the Bayes method and the cluster. Therefore, this method enhances throughput, lowers failure rates, and maximises the load balancing impact. A method \cite{50} for a load scheduling algorithm for cloud networks was put forth by A. Paulin Florence et al. The firefly method is improved in this research in terms of how it manages a set of requests and servers in the simulated cloud network. They used an index table to manage the load while maintaining the accessibility of virtual servers and requests. In that paper, the attributes have assigned to control the load and prepared the schedule list which is optimized by firefly algorithm. The Qi Liu et al. strategy for the consumer-centric perspective \cite{57} describes how to speed up the processing of incoming jobs in the heterogeneous cloud environment. 
\begin{figure}[!htbp]
	\centering
	\includegraphics[clip,trim=0cm 0cm 0cm 0cm, width=0.52\linewidth, height=9.5cm]{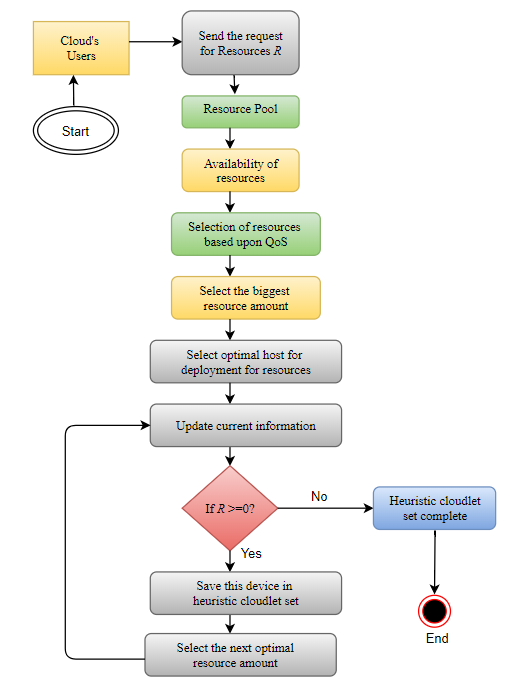}
	
	\caption{Heuristic Deployment}
	\label{Load Balancing Categorization4}
\end{figure}

Based on real-time, MapReduce was employed to optimise the execution time with the help of a prediction model. Their work shows that they allocate the tasks evenly and efficiency gets improved. There is a parallel implementation using maximum likelihood for phylogenetic analysis of DNA sequences proposed by Shamita Malik et. al. \cite{59} It distributes the cluster to all processors so that all jobs can finish on time and use their processors effectively. In order to focus on effective resource use, Shridhar et al. \cite{48} created a system that ensures the use of resources in an intelligent manner. Additionally, they were compared to earlier Active-VM algorithms.\\
\\
\textbf{Based on Multi-objective Deployment}\\
\\
The online power management system, created by Xiang Deng et al. \cite{58}, lowers the operational costs and carbon footprints of data centres. The adoption of green datacenters powered by renewable energy was advised by these authors. They recommended an online algorithm called EcoPower, which simultaneously improves load balancing and practises eco-conscious power management. Its major goal is to reduce the typical time and cost without sacrificing quality. For the creation and management of this online programme, they also used the Lyapunov optimization method. The effectiveness of this system so demonstrates that this approach successfully strikes a balance between power savings, cost, and quality. Additionally, the cost has dropped by 20\%, and it is now possible to apply complementing solar or wind energy strategies. To improve power and performance across cloud data centres, Junwei Cao et al. \cite{72} presented an efficient power allocation technique for multiple CPUs. Their method's effectiveness is reliant on a few optimization issues, such as correcting one aspect while minimising another. It is commonly known that performance improvement and power reduction are both crucial components for cloud service providers to effectively use all of the resources at their disposal. An investigation of the placement of virtual machines in IaaS clouds to increase the efficiency of data centres and cut down on energy usage was conducted in An Optimization of Virtual Machine Placement in Banker Algorithm for Energy Efficient in Cloud \cite{62}. Their findings indicate that they minimised the 770 number of VM migrations and lowered their energy consumption to 23.01 kWh with a percentage of 0.00029. The cloud-based load balancing multimedia system [83] is performed on a distributed environment in which resource manager allocates the request to the clients. Traditionally, one server handles same type of task at a time and if client requests different type of task then this service will handle at a different time. But in this paper, model runs in dynamic environment which solves with genetic algorithm with an immigrant scheme. It manages the dynamic scenario of multimedia tasks and can handle more than two different types of tasks at the same time. Tingting Wang et. al. \cite{64} also presented a task scheduling based on genetic algorithm which adopts a double-fitness adaptive job spanning time and greedy algorithm to initiate the population. It proves the more optimized and compared the performance through simulations. The authors in [85] have analysed the performance of dynamic adaptive load balancing is better than among all the algorithms. Keng-Mao Cho et. al. \cite{17} proposed the approach of load balancing in cloud computing for meta-heuristic VM scheduling. They have considered the dynamic input requests and didn't bothered about what type of tasks are running on created VMs. Basically, in this paper authors have combined two popular approaches i.e. ant colony optimization (ACO) and particle swarm optimization (PSO). Their new technique is named as ACOPS. They deal with three resources for load balancing i.e memory, cpu utilization and disk utilization.
\section{Resource Management}
Resource provisioning is an accountable for meeting user needs in accordance with quality of service (QoS) requirements, SLA agreements, and resource matching to anticipated workloads \cite{61} \cite{63} \cite{65}. Depending on the necessary quality of service (QoS) criteria, the activity that should be performed can be determined by scheduling. Scheduling is in charge of choosing the best virtual machines to use for task execution using heuristic or meta-heuristic algorithms \cite{15} \cite{16}. Additionally, it is responsible for making sure QoS requirements are met. As shown in Figure 5, Resource management is categorized as of three phases. The approach of load balancing aids in the management and efficient use of resources. Since the creation of the cloud paradigm, it has been a problem. In order to increase the number of concurrent users and overall stability of applications, load balancing is the process of reassigning the total load to each particular node \cite{66}. To improve resource utilisation and system response time, the workload is divided among two or more servers, hard discs, network interfaces, or other computing resources. Each system in the network is guaranteed to have the same amount of work at all times thanks to a load balancing approach. This results in the best response time and efficient use of the available resources \cite{19} \cite{20} and means that neither one of them is unduly overloaded nor underutilised. The cloud provider must be able to handle the increasing workload in a way that is scalable, and they must also locate the virtual machines optimally to take into account the increase in traffic for better load balancing.\\
\begin{figure} [htbp]
	\centering
	\includegraphics[clip,trim=0cm 0cm 0cm 0cm, width=0.92\linewidth, height=10.5cm]{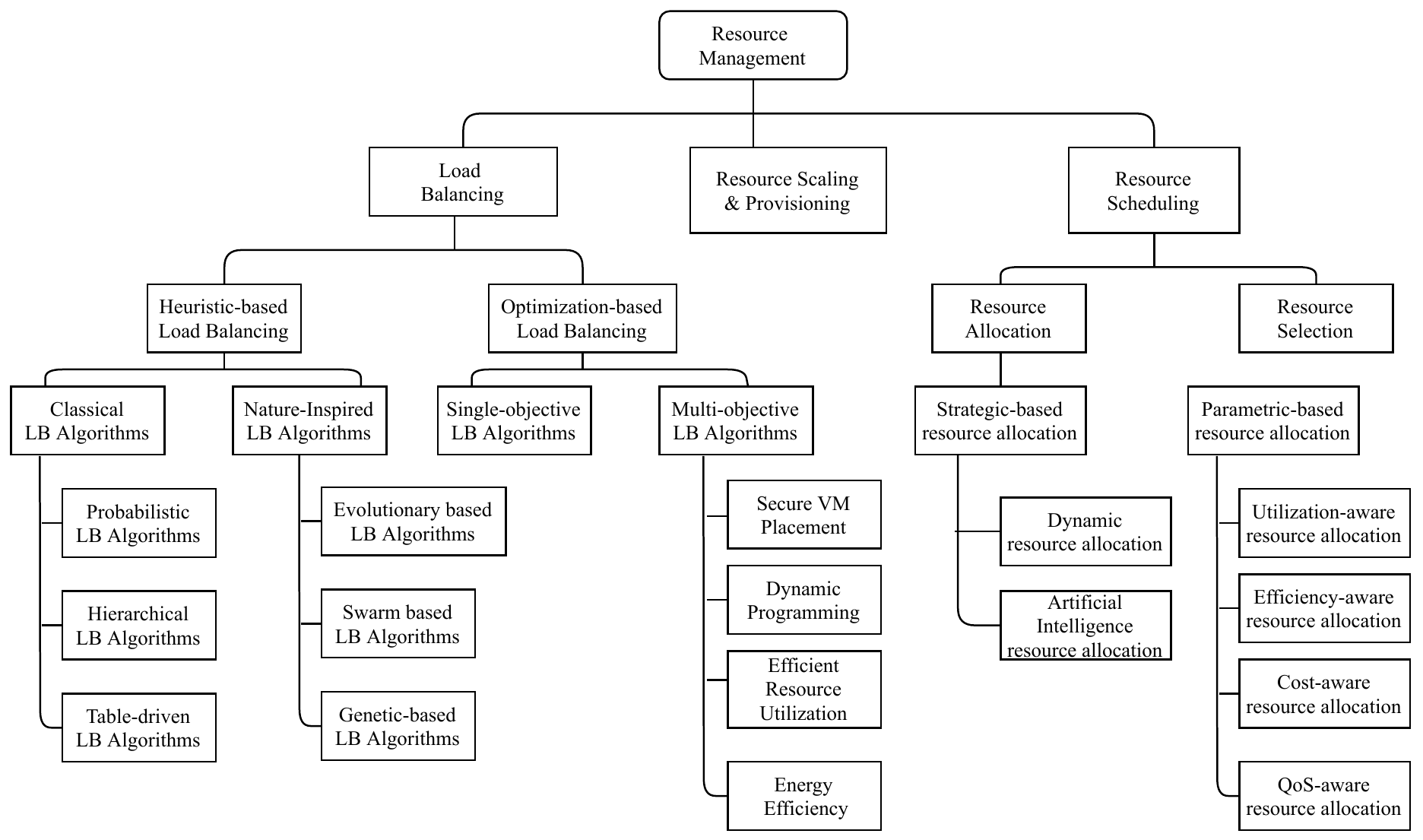}
	\caption{Resource Management Categorization}
	\label{Load Balancing Categorization5}
\end{figure}

In the networks of cloud data centres, traffic has grown tremendously due to the quick growth of applications. It becomes more frequent to relocate VMs to other hosts for better load balancing in order to meet the performance needs of virtual machines (VMs). Secure VM allocation has drawn a lot of attention as one of the most important issues in cloud computing. Users have always been at danger when sharing private information in the cloud since multiple tenants use the same resources. One such risk is a co-resident attack, in which a hostile user infiltrates all VMs running on the same server and obtains the personal data of authorised users. It goes without saying that co-resident assaults are more likely when attackers and clients share a server. If the malicious VMs can be found before they are executed, these attacks can be stopped by blocking them. But in a real-world situation, it is very difficult to accomplish. Therefore, a security-aware load balancing architecture that also ensures efficient use of computing resources and power consumption is important \cite{67} \cite{68}. There are two categories of secure resource allocation in the cloud: (1) Implementing a secure load balancing approach to map virtual machines (VMs) onto safe physical machines that are security aware VMs for task deployment. It greatly increases the challenges for attackers to obtain co-residence and (2) Workload Mapping onto VMs \cite{69} \cite{70}. It is usual practise for VMs of various users to run at the same physical server (i.e., these VMs are co-resident) and to be logically segregated from one another in order to maximise the utilisation rate of the hardware platform.
\subsection{Resource Scheduling Techniques}
The best technique to decide which activity should be carried out based on the necessary quality of service (QoS) standards is through scheduling. Scheduling is in charge of choosing the best virtual machines for a task's execution using a heuristic or meta-heuristic algorithm and is also in charge of making sure the QoS constraint is met. The goal of resource provisioning with scheduling (RPS) is to provide users with virtual machines while maintaining SLAs and meiiianticipated workload (applications). After a thorough analysis of the workload, a service level agreement (SLA) commitment is created between users and the service provider. Before workload or tasks are distributed among virtual computers (resources), cloud-running resources are monitored and their individual loads are computed. Tasks are not provided to resources of this type if any virtual machines are overutilized. The next workload is mapped with the resources that are available, and it is determined whether or not the virtual machine that is now running is sufficient to carry out the workload. Running resources are raised utilising the horizontal scaling concept if necessary; otherwise, the workload is apportioned, and the necessary QoS parameters are determined \cite{71} \cite{72}.
\begin{figure} [htbp]
	\centering
	\includegraphics[clip,trim=0cm 0cm 0cm 0cm, width=0.92\linewidth, height=14.5cm]{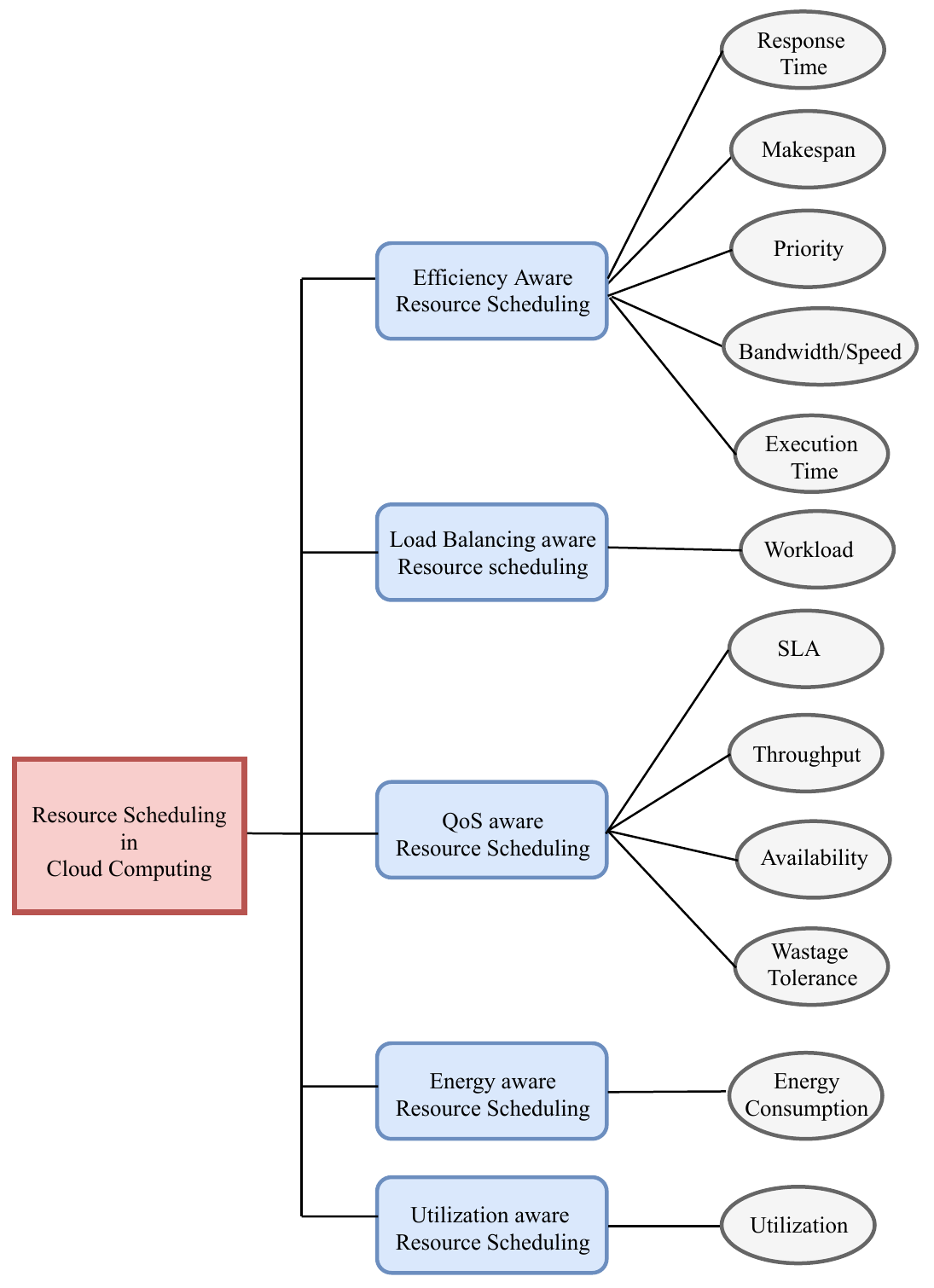}
	\caption{Resource Scheduling Techniques}
	\label{Load Balancing Categorization6}
\end{figure}
According to Figure. 6, the resource scheduling plans are divided into five hybrid types, including utilisation aware resource scheduling, efficiency aware resource scheduling, energy aware resource scheduling, load balancing aware resource scheduling, and QoS aware resource scheduling \cite{69}. This categorization is based on the metric that was utilised to assess performance during our research. The oval shapes stand in for the measures that the writers used to assess how well their suggested works performed based on these parameters. When scheduling resources with efficiency in mind, the quantity of resources used for processing is expressed based on the targeted resources. The response time, execution time, makespan, bandwidth/speed, and priority can all be improved and enhanced with the use of an efficient resource scheduling system \cite{73} \cite{74}. Through the sharing of loads across data centre infrastructures, load balancing is a workable method that enhances virtual machines and data centres overburdened with computing cloudlets, tasks, or jobs to produce a competent performance of the systems. A major problem in cloud computing is resource scheduling that is QoS aware \cite{52} \cite{59}. It entails efficiently allocating users' requested tasks to various resources in accordance with QoS, which emphasises the fault tolerance, availability, dependability, throughput, recovery time, and SLA of both cloud providers and users \cite{46}. Techniques for resource scheduling that are energy conscious are needed to solve the issues arising from the data centres' excessive energy consumption \cite{75} \cite{71}.
The performance of related frameworks are evaluated using following metrics.
\begin{itemize}
    \item \textbf{Makespan}: The experiment evaluates the makespan time, which defines the overall amount of time required to complete the activities from start to finish. When job requests are assigned to physical hosts during scheduling, it is primarily used \cite{45}. As more jobs are requested, the processing time grows. In order to optimise their financial advantage, CSPs always look for the best server and virtual machine configurations. To calculate the last task completion time \cite{26} \cite{38} \cite{40} \cite{70} on VM $j$ as in eq. 1:
    \begin{equation}
TC_j^M=\sum_{k=1}^{r-1}p_j+(k-1)*m\,\,\,,\,\,\,\,\forall j=\{1,2,\dots,m\}
\end{equation}
The optimal makespan time can be calculated as in eq. 2:
\begin{equation}
MS_{opt}=MAX_{k=1,2,\dots,m} \sum_{l=1}^{k} TC_l^M +p_{n-l+1}/k
\end{equation}
\item	\textbf{Resource Utilization}: As resource utilization is the very important parameter to measure the efficiency of any scheduling algorithm. The experiments emphasize on finding out the optimal VM which takes less amount of resources for deploying the cloudlets effciently \cite{71} \cite{72}. The average resource utilization of datacenter that needs to be maximized is demonstrated using eq. 3. |N| represents the number of resources to be considered i.e. |N|= 3.
\begin{equation}
\Phi_\delta=\frac{\sum_{i=1}^{t} \Phi_{i=1}^C + \sum_{i=1}^{t} \Phi_{i=1}^M +  \sum_{i=1}^{t} \Phi_{i=1}^B }{|N| \times \sum_{i=1}^{t} \varTheta_i}
\end{equation}
The resource utilization of each resource is calculated independently. The utilization of one physical host is computed using eq. 4, either CPU or memory or bandwidth respectively \cite{66}.
\begin{equation}
\Phi_i^R= \frac{\sum_{i=1}^{n} \varTheta_i \times {VM}_j^R }{\uplus_i^R}\,\, R\in C|M|B
\end{equation}
\item	\textbf{Energy Efficiency}: In cloud data centers, the energy consumption has dramatically increased \cite{73}. The objective of our thesis is to maximize the resource utilization as well as minimize the consumption of energy. It says that energy consumption has a linear relationship with processor utilization. It proves that CPU consumption is the main part of energy. The authors have assumed that SMs in cloud DCs work in three different states: active state, idle state and sleep state. In order to minimize the power consumption, idle servers will be switched off to sleep mode because an idle server consumes about 70\% of total power consumption as in eq. 5.
\begin{align}
\exists_k=
\begin{cases}
\Im \times \Big[\Phi_i^R \times \eth_k^{active} + (1-\Phi_i^R) \times \eth_k^{idle}\Big],Host state:on  \\
\Im \times \eth_k^{sleep},\,\,\,\,\, \, \, \, \, \, \, \,\,\,\, \, \,Host-state:\, off
\end{cases}
\end{align}
The total energy consumption of a cloud data center is defined as $\exists=\sum_{k=0}^{n}\exists_k$. If the host state is ON then it comes either in active state or in idle state.\\
\end{itemize}

\section{Emerging Challenges and Future Research Directions}
Being a trending strategy today, cloud has garnered interest as a way to lower capital expenditure and boost system effectiveness. It keeps user data on a sizable virtual storage system made up of several servers connected by a network. There have been numerous ways and solutions put out to prevent data loss or leakage as well as load management, however the solutions still have certain research gaps.
We require a perfect solution that can manage secret data, safeguard it, manage the load as well as resources wisely and find the bad actors and their activities. Using the literature evaluation as a guide, the following research holes are discovered:
\begin{itemize}
    \item Data is hosted on un-trusted servers: The cloud users will need to ensure that their data remains confidential and transparent only to authorized users. If the company can't guarantee that its data is kept confidential, it might not be worth using Cloud Computing to store the information. This could lead to loss of customers' trust, fines from a regulator, and even financial losses.\\
    \item Single-objective approaches: The approaches provide a solution for manage the load maximum utilization of resources. These approaches work well for data that is stored in one place or that is used in one way. Some useful strategies are effective but do not recognize data as required. To keep data safe as well as balance, we need a hybrid approach that meets all our needs.\\
    \item Static requests: Numerous hosts and application requests with changing resource requirements are present in the cloud datacentre. Diverse demands are made on the distribution of resources. These variables may result in load imbalances, which have an impact on resource usage and scheduling effectiveness. Therefore, an effective load balancing models for resource allocation is essential to bring the operational efficiency with improved scalability. 
    \item Service Quality Assurance: It is the process of ensuring that the quality and required services are delivered in a cloud service model. The document calls for using the best cloud service model possible, including public, private, hybrid, and community clouds.
    \item Secure VM allocation: The secure VM placement is another aspect that is explored to achieve security in the context of load balancing. By examining ways to enhance the VM allocation strategy, this work addresses the issue from a different angle, making it more difficult for attackers to maintain co-tenancy with their targets.
\end{itemize}

The following are the research directions in order to fill the identified research challenges:
\begin{itemize}
    \item [$\checkmark$] Focusing on derive a new technique on how to balance the upcoming load in cloud.
    \item [$\checkmark$] How to choose the most optimal host for deploying the tasks.
     \item [$\checkmark$] 	Provide secure multi-owner data sharing in dynamic groups.
      \item [$\checkmark$] To increase the performance significantly. 
       \item [$\checkmark$] Based on security analysis of VM allocation policies, able to prevent the attackers in the
          mitigation of threats. 
        \item [$\checkmark$] Efforts will be made to create an efficient technique for achieving both security and efficiency for  
           the virtual machine allocation policies.
\end{itemize}
\section{Conclusion and Future scope}
As these new cloud-based solutions are developed, they first identify the security and privacy standards that are currently in place and address the issues that consumers are having with regard to quality, efficiency, and maximizing the satisfaction of their needs. In order to validate our approaches, we fully understand the importance of safe cloud computing and resource efficiency. Because security aids in preventing data leaks and disposals, load balancing aids in improving workload distribution among several computing resources. The proposed or what we are considering planning contributes to:
\begin{itemize}
    \item [$\checkmark$] Machine learning algorithm can be used to predict the upcoming data rate or application requests for better scalability in cloud computing.
    \item [$\checkmark$] Some of the models are heuristic optimization models that can be extended by employing meta-heuristic optimization.
    \item [$\checkmark$] The proposed algorithm can be used to improve the others QoS parameters like reliability, elasticity, mean time to failure, prediction, security, performance, SLA violation and response time for better cloud services.
    \item [$\checkmark$] Developed LB algorithms can be tested in future at Montage, EpiGenomics, CyberShake, LIGO, SIPHT realistic workflows.
    \item [$\checkmark$] The load balancing models can be developed using other learning paradigms including deep learning, quantum inspired learning, capsule networks etc.
    \item [$\checkmark$] Decision makers in selecting the optimal cloud service provisioning.
    \item [$\checkmark$] Practically used in the cloud for secure data sharing among the group enhances trust level in cryptographic services \cite{1}.
\end{itemize}

  

\end{document}